\documentclass{iau} 
\usepackage{graphicx}
\usepackage{natbib}

\pubyear{2022}
\volume{370}
\jname{Winds of Stars and Exoplanets}
\editors{A.~A. Vidotto, L.~Fossati \& J.~Vink, eds.}

\title[Winds in RSGs: Interferometry] 
{The effect of winds in red supergiants: modeling for interferometry}

\author[Gonz\'alez-Tor\`a et al]   
{Gemma Gonz\'alez-Tor\`a$^{1,2}$, Markus Wittkowski$^1$, Ben Davies$^2$ \and Bertrand Plez$^3$}

\affiliation{$^1$ European Southern Observatory (ESO), Karl Schwarzschildstrasse 2, 85748 Garching bei M\"unchen, Germany
 \\ [\affilskip]
$^2$ Astrophysics Research Institute, Liverpool John Moores University, 146 Brownlow Hill, Liverpool L3 5RF, United Kingdom
\\ [\affilskip]
$^3$ LUPM, Universit\'e de Montpellier, CNRS, 34095 Montpellier, France
 \\ [\affilskip]
}

\begin{document}

\maketitle

\begin{abstract}
Red supergiants (RSGs) are evolved massive stars in a stage preceding core-collapse supernova. Understanding evolved-phases of these cool stars is key to understanding the cosmic matter cycle of our Universe, since they enrich the cosmos with newly formed elements. However, the physical processes that trigger mass loss in their atmospheres are still not fully understood, and remain one of the key questions in stellar astrophysics. We use a new method to study the extended atmospheres of these cold stars, exploring the effect of a stellar wind for both a simple radiative equilibrium model and a semi-empirical model that accounts for a chromospheric temperature structure. We then can compute the intensities, fluxes and visibilities matching the observations for the different instruments at the Very Large Telescope Interferometer (VLTI). Specifically, when comparing with the atmospheric structure of HD~95687 based on published VLTI/AMBER  data, we find that our model can accurately match these observations in the $K$-band, showing the enormous potential of this methodology to reproduce extended atmospheres of RSGs.
\end{abstract}

\firstsection 
              
\section{Introduction}\label{sec:intro}
When evolved massive stars leave the main sequence and start the red supergiant (RSG) phase, stellar winds can impact the final fate in their evolutionary path \citep{1986ARA&A..24..329C}. These mass-loss events are initiated in the extended atmospheres of RSGs: part of their material is ejected and transported up to several radii. Beyond that, the temperature is cold enough to start condensing the ejected material into dust grains. 

However, the mechanism that triggers these mass-loss events in the extended atmospheres of RSGs is still poorly understood. There have been some attempts to explain the mechanism of stellar winds in RSGs \citep[e.g.,][]{2000ARA&A..38..613K,2007A&A...469..671J,2021A&A...646A.180K}, but there is still no consensus. 


The current models use the following atmospheric structure: first a stellar modelization up to the photosphere (defined where $\tau_{\mathrm{Ross}}=2/3$) with model atmosphere grids such as {\sc MARCS} \citep{2008A&A...486..951G}, and then adding the contribution of dust modeling, such as DUSTY \citep{1997MNRAS.287..799I}. The atmospheric extension from the photosphere to where the dust is formed is usually left empty, since we do not know the physical processes that trigger these mass-loss events. 


In this work, we aim to "fill" the atmospheric extension from the photosphere to the dusty shell. We use the model developed by \citet{2021MNRAS.508.5757D}, where they explored the extension of the atmospheres close to the stellar surface at radii smaller than the inner dust shells, in the optical and near-IR, by adding the influence of a stellar wind in the {\sc MARCS} model atmospheres.

For the purpose to study the extension of the atmosphere, we use interferometric data. Interferometry uses an array of telescopes to increase the angular resolution of the observations. By using interferometry, we have high-spatial resolution data of the stellar atmospheres of RSGs. This is an additional information that spectroscopy does not fully provide. Therefore, it is a very powerful tool to study the structure of extended atmospheres in detail.

\section{Model}\label{sec:model}
Our model is based on \citet{2021MNRAS.508.5757D}, where they start with a {\sc MARCS} model atmosphere and plug-in the effect of a stellar wind. This model assumes local thermodynamic equilibrium (LTE), hydrostatic equilibrium, and spherical symmetry. We extend the model with a radius stratification up to $\sim8.5\,R_{\star}$, where $R_{\star}$ is defined as the radius where the Rosseland opacity $\tau_{\mathrm{Ross}}=2/3$. For a detailed discussion about the limitations and assumptions, the reader is referred to \citet{2021MNRAS.508.5757D}.

To determine the wind density, we use the mass continuity expression,
\begin{equation}\label{eq:masscont}
\dot{M}=4\pi r^{2}\rho(r)v(r)
\end{equation}
where $\rho$ and $v$ are the density and velocity as a function of the stellar radial coordinate $r$ respectively. The wind density $\rho_{\mathrm{wind}}(r)$ has the shape proposed by \citet{2001ApJ...551.1073H}: 
\begin{equation}\label{eq:betalaw}
\rho_{\mathrm{wind}}=\frac{\rho_{\mathrm{phot.}}}{(R_{\mathrm{max}}/R{\star})^{2}} \left( 1-\left( \frac{0.998}{(R_{\mathrm{max}}/R_{\star})} \right) ^{\gamma} \right) ^{\beta}
\end{equation}
where $R_{\mathrm{max}}$ is the arbitrary outer-most radius of the model, in our case $8.5\,R_{\star}$. The $\beta$ and $\gamma$ parameters define the smoothness of the extended wind region and were initially set in the semi-empirical 1D model of $\alpha$ Ori by \citet{2001ApJ...551.1073H}: $\beta_{\mathrm{Harp}}=-1.10$ and $\gamma_{\mathrm{Harp}}=0.45$. 

The velocity profile is found assuming a fiducial wind limit of $v_{\infty}=25\pm5$ km/s, that is the value matched to \citet{1998IrAJ...25....7R,2005A&A...438..273V}, and Equation~\ref{eq:masscont}.

For the temperature profile we first used simple radiative transfer equilibrium (R.E.), that will result in a smoothly decreasing temperature profile for the extended atmosphere. We also defined a different temperature profile based on spatially-resolved radio continuum data of $\alpha$~Ori by \citet{2001ApJ...551.1073H}. The main characteristic of this profile is a temperature inversion in the chromosphere of the star, that peaks at $\sim 1.4\,R_{\star}$, and decreases again.

\begin{figure}[ht]
\begin{center}
 \includegraphics[width=0.6\textwidth]{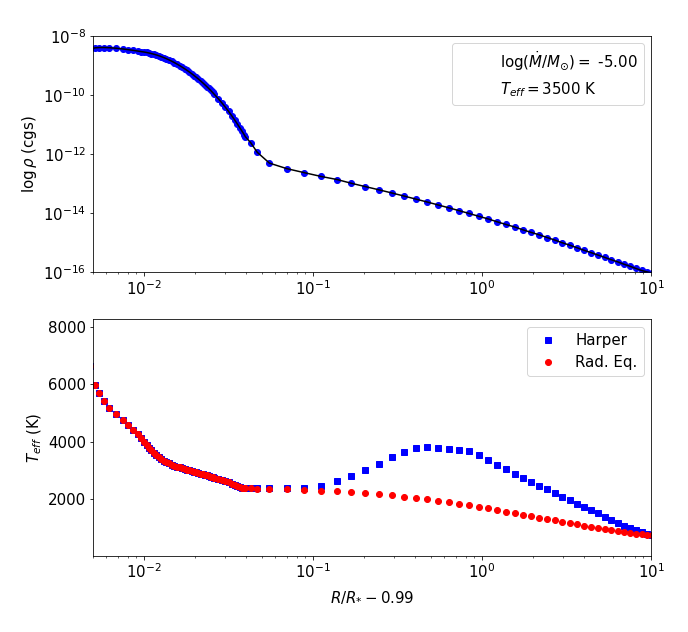} 
 \caption{Density and temperature profiles (Harper temperature inversion in blue, simple R.E. in red) for the extended model. } \label{fig:model}
\end{center}
\end{figure}

Figure~\ref{fig:model} shows the the density (upper panel) and the temperature structures (lower panel) of our model, the latest for simple R.E. (blue squares) and a temperature chromospheric inversion (red circles).

\section{Results}\label{sec:results}
\subsection{For base model}

We compute the spectra, intensities and squared visibility amplitudes ($|V|^{2}$) for a base model of $T_{\mathrm{eff}}=3500$ K, $\log g=0.0$, $[Z]=0$, $\xi=5$ km/s, $M=15\,M_{\odot}$, $R_{\star}=690\,R_{\odot}$ and $R_{\mathrm{max}}=8.5\,R_{\star}$, corresponding to a RSG similar to HD~95687 \citep{2015A&A...575A..50A}. The density parameters in Equation~\ref{eq:betalaw} are $\beta_{\mathrm{Harp}}=-1.10$ and $\gamma_{\mathrm{Harp}}=0.45$ as in \citet{2001ApJ...551.1073H} and the wind limit $v_{\infty}=25$ km/s. The temperature profile is initially set to simple R.E.. We use mass-loss rates of $\dot{M}=10^{-4}$, $10^{-5}$, $10^{-6}$ and $10^{-7}$ $M_{\odot}/\mathrm{yr}$, and a simple {\sc MARCS} model without any wind. As an example, we simulate a star with an angular diameter at the photosphere of $\theta_{\mathrm{Ross}}=3$ mas, a baseline of $B=60$ m and without any additional over-resolved component. 

\begin{figure}[ht]
\begin{center}
 \includegraphics[width=1.\textwidth]{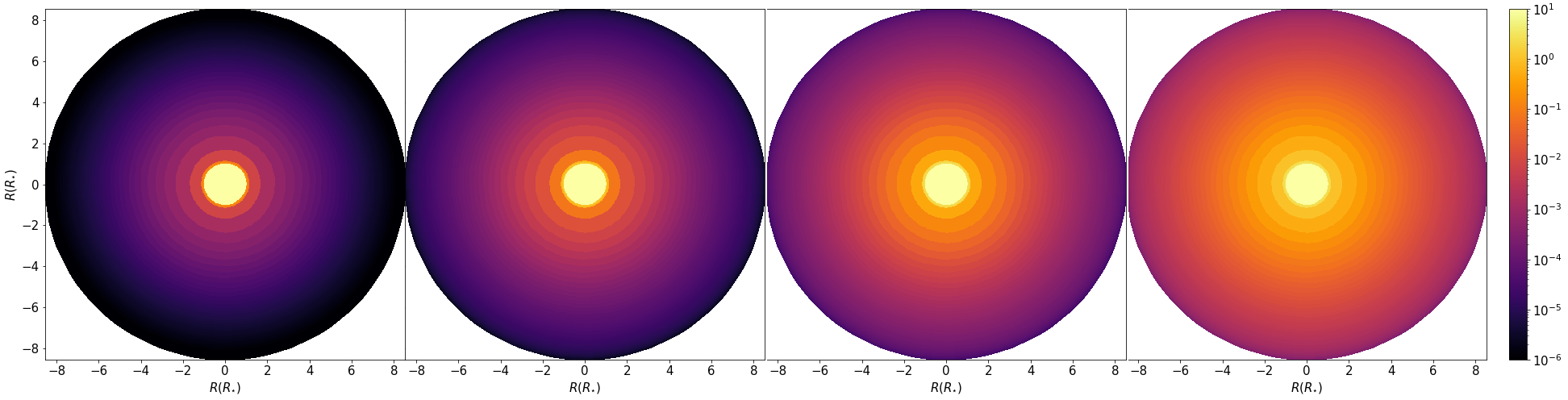} 
 \caption{2D profile of our atmospheric model, from left to right: $\dot{M}=10^{-7}$, $10^{-6}$, $10^{-5}$ and $10^{-4}$ $M_{\odot}/\mathrm{yr}$. This is the case of simple R.E..} \label{fig:winds}
\end{center}
\end{figure}

Figure~\ref{fig:winds} shows the 2D intensities for mass-loss rates of  $\dot{M}=10^{-7}$, $10^{-6}$, $10^{-5}$ and $10^{-4}$ $M_{\odot}/\mathrm{yr}$ (panels from left to right), for a star with an extended radius of $R_{\mathrm{max}}=8.5\,R_{\star}$ for a cut in the transition {\sc CO (2-0)} ($\lambda=2.29\,\mathrm{\mu m}$). We see that, unlike MARCS, all models show extension at $R>1\,R_{\star}$. As expected, as we increase the mass-loss rate we also have more extension. 

\begin{figure}[ht]
\begin{center}
 \includegraphics[width=0.7\textwidth]{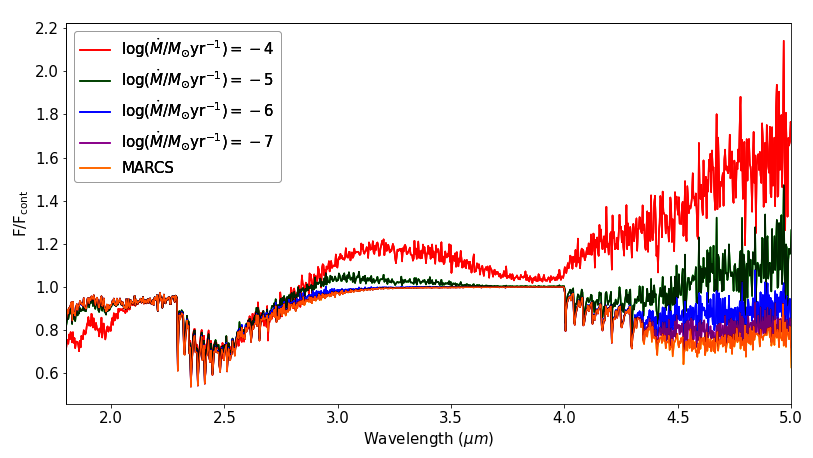} 
 \caption{Normalized flux for the different $\dot{M}$. This is the case of simple R.E..} \label{fig:flux}
\end{center}
\end{figure}

Figure~\ref{fig:flux} shows the flux for a wavelength range of $1.8\,\mathrm{\mu m}<\lambda<5.0\,\mathrm{\mu m}$ corresponding to the $K$, $L$ and $M-$bands, for simple MARCS model (orange), $\dot{M}=10^{-7}$ (purple), $10^{-6}$ (blue), $10^{-5}$ (dark green) and $10^{-4}$ $M_{\odot}/\mathrm{yr}$ (red). Compared to a MARCS model, as we increase the mass-loss rate we start seeing more features in the spectra (e.g., the water absorption in  $1.8\,\mathrm{\mu m}<\lambda<2.0\,\mathrm{\mu m}$ and $3.0\,\mathrm{\mu m}<\lambda<3.5\,\mathrm{\mu m}$, the SiO and CO emissions in $\lambda>4.0\,\mathrm{\mu m}$). 

\begin{figure}[ht]
\begin{center}
 \includegraphics[width=0.7\textwidth]{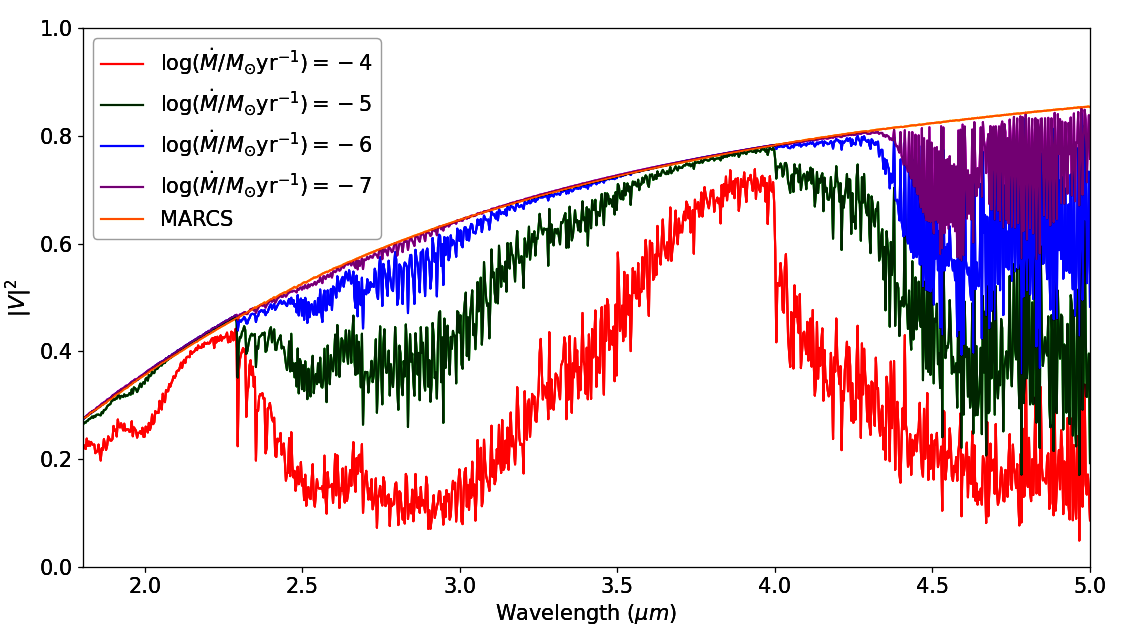} 
 \caption{Same as Figure~\ref{fig:flux} but for the $|V|^{2}$.} \label{fig:v2}
\end{center}
\end{figure}

To calculate the squared $|V|^{2}$ we used the Hankel transform as in \citet{2000MNRAS.318..387D}. Figure~\ref{fig:v2} shows the same as Figure~\ref{fig:flux} but for $|V|^{2}$. We the same features as Figure~\ref{fig:flux} (e.g., the water absorption in  $1.8\,\mathrm{\mu m}<\lambda<2.0\,\mathrm{\mu m}$ and $3.0\,\mathrm{\mu m}<\lambda<3.5\,\mathrm{\mu m}$, the SiO and CO emissions in $\lambda>4.0\,\mathrm{\mu m}$), and the extra CO extension in $2.3\, \mathrm{\mu m}<\lambda<3.1\,\mathrm{\mu m}$. These features show that our model is able to extend the atmosphere, unlike simple MARCS which has no features in the $|V|^{2}$. 

\subsection{Deviations from base model}\label{sub:dev}
We can change the $\beta$ and $\gamma$ parameters of Equation~\ref{eq:betalaw} that define the density profile. This will affect the wind density profile close to the photosphere, as we decrease the value of the parameters the density becomes steeper in that region. This will have an effect in the lines formed close to the stellar surface \citep[e.g., water layers][]{2020A&A...642A.235K}, and therefore are sensitive to variations of the density profile in this region. 

As for the temperature profile, the main difference between R.E. and the chromospheric temperature inversion is seen in the CO for the $K$-band. Even for lower mass-loss rates $\log \dot{M}/M_{\odot}<-5$ we observe the CO lines very depleted in the flux when using the temperature inversion profile, and already in emission for $\log \dot{M}/M_{\odot}\sim-4$. This is not the case of R.E., since the CO is always in absorption in that region (Figure~\ref{fig:flux}).  When compared to the observations, we always see the CO lines in absorption, confirming that R.E. may be more appropriate to fit the observational data. Depending on the wavelength range of observations, the temperature peak at the chromosphere varies: CO MOLsphere data derived a temperature of $\sim 2000$ K at $1.2-1.4\,R_{\star}$ \citep{1998Natur.392..575L}, while for the optical and ultraviolet the peak temperature is $\sim5000$ K at similar radii \citep{2013A&A...555A..24O}. 
\citet{2020A&A...638A..65O} suggested these components co-exist in different structures at similar radii in an inhomogeneous atmosphere. Observations at different wavelengths may then be sensitive to different such structures.


\subsection{Case study}
We compare our model to published VLTI/AMBER data of the RSG HD~95687  available by \citet{2015A&A...575A..50A}.The data were taken using the AMBER medium-resolution mode ($R \sim 1500$) in the $K-2.1\,\mathrm{\mu m}$ and $K-2.3\,\mathrm{\mu m}$ bands. 

For our model fit, we check both the spectra and $|V|^{2}$. We use a range of mass-loss rates of $-7<\log \dot{M}/M_{\odot}<-4$ with a grid spacing of $\Delta\dot{M}/M_{\odot}=0.25$, and density parameters $-1.1<\beta<-1.60$ in steps of $\Delta\beta=0.25$ and $0.05<\gamma<0.45$ in steps of $\Delta\gamma=0.2$. For the temperature profile we use R.E., since the temperature inversion would either show depleted CO lines ($\lambda=2.5-3\,\mathrm{\mu m}$) for the spectra, which do not match with the observations, or not enough extension for the $|V|^{2}$. Therefore it is not possible to find a model with the temperature inversion profile that fits both spectra and $|V|^{2}$ simultaneously (see Section~\ref{sub:dev}).

\begin{figure}[ht]
\begin{center}
 \includegraphics[width=1.\textwidth]{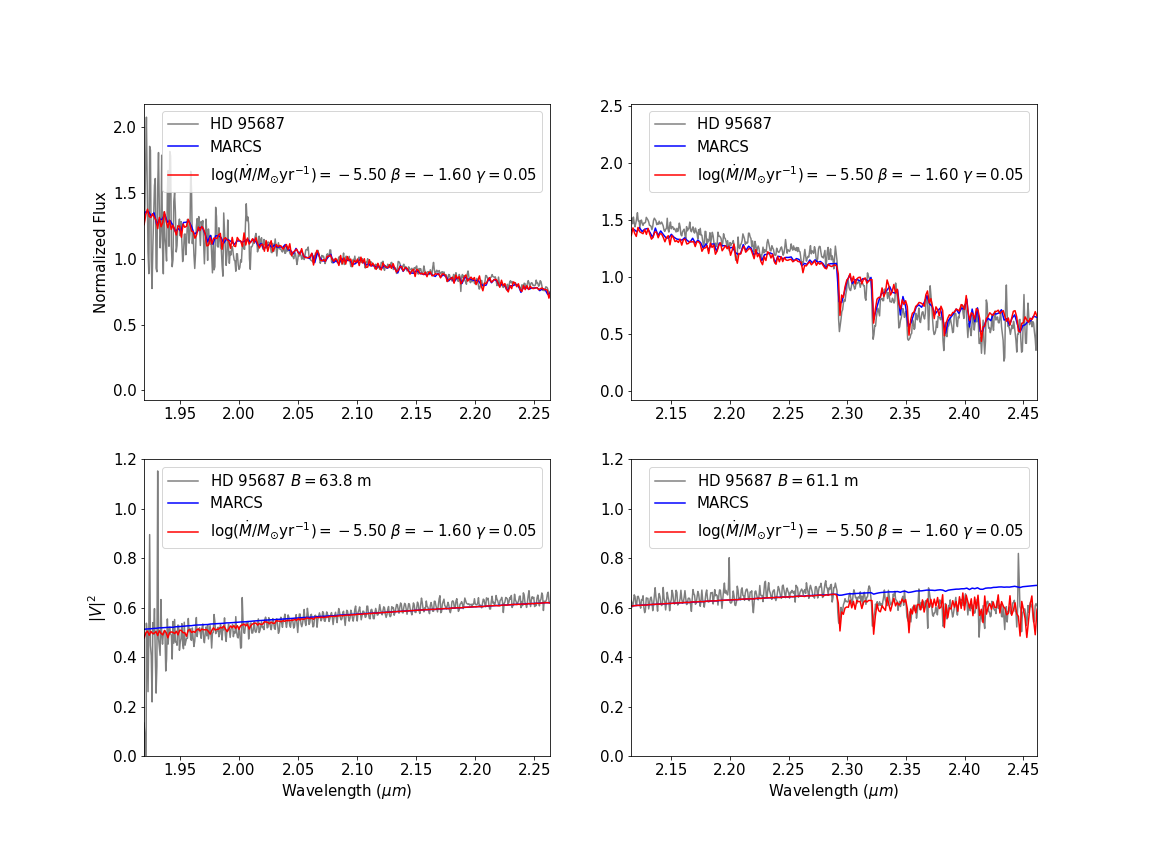} 
 \caption{Normalized flux and $|V|^{2}$ for HD 95687 as observed with VLTI/AMBER (in gray), the best fitting results for this work in red, and the best MARCS fitting in blue.} \label{fig:c1}
\end{center}
\end{figure}

Figure~\ref{fig:c1} shows the {\sc MARCS} model fit to the data of \citet{2015A&A...575A..50A} (blue), our best {\sc MARCS}+wind model fit with $\log \dot{M}/M_{\odot}=-5.50$, $\beta_{\mathrm{Harp}}=-1.60$ and $\gamma_{\mathrm{Harp}}=0.05$ (red), compared to the data of HD~95687 (gray). We see that our model can fit well both flux and $|V|^{2}$. There is a big improvement with respect to MARCS for the $|V|^{2}$, especially in the region where the CO is present: $2.3\, \mathrm{\mu m}<\lambda<3.1\,\mathrm{\mu m}$.  In addition, the density profile is steeper than expected, but the $\dot{M}$ is reasonable when compared with typical mass-loss prescriptions \citep[e.g.,][]{1988A&AS...72..259D,2005ApJ...630L..73S,2020MNRAS.492.5994B}.


\section{Conclusions}\label{sec:conc}
This is the first extended atmosphere model to our knowledge that can reproduce in great detail both the spectra and $|V|^{2}$ simultaneously. Therefore, we have shown the immense potential of this semi-empirical model of {\sc MARCS}+wind, not only to match the spectral features without the need of dust, but also the visibilities obtained by interferometric means. 

To fit both the water and CO extensions, the density shape should be steeper close to the surface of the star than previously expected by \citet{2001ApJ...551.1073H}. Regarding the temperature profile, we find that the R.E. reproduces the spectra better than the chromospheric temperature inversion, since we do not observe any emission in the CO bands. The possible reason that R.E. fits better than the temperature inversion could be the presence of different spatial cells with different temperatures in the hot luke-warm chromospheres of RSGs \citep{2020A&A...638A..65O}.

In the future, we want to compare this model with more wavelength ranges, to see the effects in different wavelengths such as the $L$ or $M$-bands. 

\section{Q\&A}\label{sec:q}

{\sc Question:} Why are there 4 panels in Figure~\ref{fig:c1} instead of 2? It seems like the wavelength regions overlap.

{\sc Answer:} This is because the visibilities were taken with different array configurations and therefore different baselines. We decided to separate both the fluxes and visibilities following \citet{2015A&A...575A..50A} to avoid confusion.




\def\apj{{ApJ}}    
\def\nat{{Nature}}    
\def\jgr{{JGR}}    
\def\apjl{{ApJ Letters}}    
\def\aap{{A\&A}}   
\def\mnras{{MNRAS}}
\def\aj{{AJ}}
\let\mnrasl=\mnras


\bibliographystyle{aa}
\bibliography{biblio.bib}

\begin{thebibliography}{19}
\expandafter\ifx\csname natexlab\endcsname\relax\def\natexlab#1{#1}\fi

\bibitem[{{Arroyo-Torres} {et~al.}(2015){Arroyo-Torres}, {Wittkowski},
  {Chiavassa}, {Scholz}, {Freytag}, {Marcaide}, {Hauschildt}, {Wood}, \&
  {Abellan}}]{2015A&A...575A..50A}
{Arroyo-Torres}, B., {Wittkowski}, M., {Chiavassa}, A., {et~al.} 2015, \aap,
  575, A50

\bibitem[{{Beasor} {et~al.}(2020){Beasor}, {Davies}, {Smith}, {van Loon},
  {Gehrz}, \& {Figer}}]{2020MNRAS.492.5994B}
{Beasor}, E.~R., {Davies}, B., {Smith}, N., {et~al.} 2020, \mnras, 492, 5994

\bibitem[{{Chiosi} \& {Maeder}(1986)}]{1986ARA&A..24..329C}
{Chiosi}, C. \& {Maeder}, A. 1986, \araa, 24, 329

\bibitem[{{Davies} \& {Plez}(2021)}]{2021MNRAS.508.5757D}
{Davies}, B. \& {Plez}, B. 2021, \mnras, 508, 5757

\bibitem[{{Davis} {et~al.}(2000){Davis}, {Tango}, \&
  {Booth}}]{2000MNRAS.318..387D}
{Davis}, J., {Tango}, W.~J., \& {Booth}, A.~J. 2000, \mnras, 318, 387

\bibitem[{{de Jager} {et~al.}(1988){de Jager}, {Nieuwenhuijzen}, \& {van der
  Hucht}}]{1988A&AS...72..259D}
{de Jager}, C., {Nieuwenhuijzen}, H., \& {van der Hucht}, K.~A. 1988, \aaps,
  72, 259

\bibitem[{{Gustafsson} {et~al.}(2008){Gustafsson}, {Edvardsson}, {Eriksson},
  {J{\o}rgensen}, {Nordlund}, \& {Plez}}]{2008A&A...486..951G}
{Gustafsson}, B., {Edvardsson}, B., {Eriksson}, K., {et~al.} 2008, \aap, 486,
  951

\bibitem[{{Harper} {et~al.}(2001){Harper}, {Brown}, \&
  {Lim}}]{2001ApJ...551.1073H}
{Harper}, G.~M., {Brown}, A., \& {Lim}, J. 2001, \apj, 551, 1073

\bibitem[{{Ivezic} \& {Elitzur}(1997)}]{1997MNRAS.287..799I}
{Ivezic}, Z. \& {Elitzur}, M. 1997, \mnras, 287, 799

\bibitem[{{Josselin} \& {Plez}(2007)}]{2007A&A...469..671J}
{Josselin}, E. \& {Plez}, B. 2007, \aap, 469, 671

\bibitem[{{Kee} {et~al.}(2021){Kee}, {Sundqvist}, {Decin}, {de Koter}, \&
  {Sana}}]{2021A&A...646A.180K}
{Kee}, N.~D., {Sundqvist}, J.~O., {Decin}, L., {de Koter}, A., \& {Sana}, H.
  2021, \aap, 646, A180

\bibitem[{{Kravchenko} {et~al.}(2020){Kravchenko}, {Wittkowski}, {Jorissen},
  {Chiavassa}, {Van Eck}, {Anderson}, {Freytag}, \&
  {K{\"a}ufl}}]{2020A&A...642A.235K}
{Kravchenko}, K., {Wittkowski}, M., {Jorissen}, A., {et~al.} 2020, \aap, 642,
  A235

\bibitem[{{Kudritzki} \& {Puls}(2000)}]{2000ARA&A..38..613K}
{Kudritzki}, R.-P. \& {Puls}, J. 2000, \araa, 38, 613

\bibitem[{{Lim} {et~al.}(1998){Lim}, {Carilli}, {White}, {Beasley}, \&
  {Marson}}]{1998Natur.392..575L}
{Lim}, J., {Carilli}, C.~L., {White}, S.~M., {Beasley}, A.~J., \& {Marson},
  R.~G. 1998, \nat, 392, 575

\bibitem[{{O'Gorman} {et~al.}(2020){O'Gorman}, {Harper}, {Ohnaka},
  {Feeney-Johansson}, {Wilkeneit-Braun}, {Brown}, {Guinan}, {Lim}, {Richards},
  {Ryde}, \& {Vlemmings}}]{2020A&A...638A..65O}
{O'Gorman}, E., {Harper}, G.~M., {Ohnaka}, K., {et~al.} 2020, \aap, 638, A65

\bibitem[{{Ohnaka} {et~al.}(2013){Ohnaka}, {Hofmann}, {Schertl}, {Weigelt},
  {Baffa}, {Chelli}, {Petrov}, \& {Robbe-Dubois}}]{2013A&A...555A..24O}
{Ohnaka}, K., {Hofmann}, K.~H., {Schertl}, D., {et~al.} 2013, \aap, 555, A24

\bibitem[{{Richards} \& {Yates}(1998)}]{1998IrAJ...25....7R}
{Richards}, A.~M.~S. \& {Yates}, J.~A. 1998, Irish Astronomical Journal, 25, 7

\bibitem[{{Schr{\"o}der} \& {Cuntz}(2005)}]{2005ApJ...630L..73S}
{Schr{\"o}der}, K.~P. \& {Cuntz}, M. 2005, \apjl, 630, L73

\bibitem[{{van Loon} {et~al.}(2005){van Loon}, {Cioni}, {Zijlstra}, \&
  {Loup}}]{2005A&A...438..273V}
{van Loon}, J.~T., {Cioni}, M. R.~L., {Zijlstra}, A.~A., \& {Loup}, C. 2005,
  \aap, 438, 273

\end{thebibliography}

\end{document}